\documentclass[twocolumn,notitlepage,superscriptaddress,nofootinbib]{revtex4-1}
\usepackage{mathtools,amssymb,bm,graphicx,hyperref}
\allowdisplaybreaks[1]

\newcommand\grad{\bm\nabla}
\newcommand\+{\dagger}
\newcommand\<{\langle}
\renewcommand\>{\rangle}
\newcommand\up{\uparrow}
\newcommand\down{\downarrow}
\newcommand\eps{\epsilon}
\newcommand\eF{\eps_F}
\newcommand\kF{k_F}
\newcommand\vF{v_F}

\newcommand\p{{\bm{p}}}
\newcommand\q{{\bm{q}}}
\newcommand\x{{\bm{x}}}
\newcommand\y{{\bm{y}}}
\newcommand\ep{\eps_\p}
\newcommand\eq{\eps_\q}
\renewcommand\H{\mathcal{H}}
\newcommand\J{\mathcal{J}}
\newcommand\N{\mathcal{N}}
\renewcommand\O{\mathcal{O}}
\newcommand\T{\mathcal{T}}
\newcommand\V{\mathcal{V}}
\newcommand\Y{\mathcal{Y}}

\begin{document}

\title{Simulating quantum transport with ultracold atoms and interaction effects}

\author{Sho Nakada}
\affiliation{Department of Physics, Tokyo Institute of Technology,
Ookayama, Meguro, Tokyo 152-8551, Japan}
\author{Shun Uchino}
\affiliation{Waseda Institute for Advanced Study, Waseda University,
Shinjuku, Tokyo 169-0051, Japan}
\author{Yusuke Nishida}
\affiliation{Department of Physics, Tokyo Institute of Technology,
Ookayama, Meguro, Tokyo 152-8551, Japan}

\date{June 2020}

\begin{abstract}
Quantum transport can be simulated with ultracold atoms by employing spin superpositions of fermions interacting with spin-dependent potentials.
Here we first extend this scheme to an arbitrary number of spin components so as to allow simulating transport through a multiterminal quantum dot and derive a current formula in terms of a spin rotation matrix and potential phase shifts.
We then show that a Fano resonance manifests itself in measuring a linear conductance at zero temperature in the case of two spin components.
We also study how a weak interparticle interaction in bulk affects quantum transport in one dimension with the bosonization and renormalization techniques.
In particular, we find that the conductance vanishes for an attractive interaction due to a bulk spin gap, while it is enhanced for a repulsive interaction by a power law with lowering the temperature or the chemical potential difference.
\end{abstract}

\maketitle

\section{Introduction}
Quantum simulation, studying conventionally inaccessible quantum problems with controllable quantum systems~\cite{Georgescu:2014}, has been one of the mainstreams in ultracold atom physics~\cite{Bloch:2012,Gross:2017}.
As a celebrated example, the Fermi-Hubbard model has been studied by loading fermionic atoms onto an optical lattice with a tunable interparticle interaction, which promises to provide insights into high-temperature superconductors in condensed matter physics~\cite{Tarruell:2018}.
More recently, considerable interest has been devoted to simulating lattice gauge theories underlying elementary particle physics~\cite{Wiese:2013,Zohar:2016}.

The spectrum of quantum phenomena that can be simulated with ultracold atoms has been significantly broadened by the idea of synthetic dimensions, which regards internal degrees of freedom such as spins as spatial degrees of freedom~\cite{Boada:2012,Celi:2014}.
Consequently, quantum Hall physics in two dimensions was successfully simulated with multicomponent atoms in a one-dimensional optical lattice~\cite{Mancini:2015,Stuhl:2015} and its extensions even toward four and higher dimensions were studied theoretically~\cite{Price:2015,Lee:2018,Petrides:2018,Ozawa:2019}.

Another application of regarding spins as spatial degrees of freedom may be mesoscopic quantum transport proposed in Ref.~\cite{Knap:2012} (see Sec.~V therein).
Here two spin superposition states of fermions play the role of ``left'' and ``right'' leads in quantum dot experiments and their interactions with spin-dependent potentials cause transports of particle numbers between the two degrees of freedom.
This scheme allows us to study the nonequilibrium orthogonality catastrophe and full-counting statistics~\cite{You:2019}, which is challenging in condensed matter experiments.
The same scheme was also adopted as a transport measurement to probe the orbital Kondo effect realized with ultracold atoms~\cite{Nishida:2013,Nishida:2016}.

The purpose of this paper is to extend the above idea of simulating quantum transport with ultracold atoms toward two directions.
One direction studied in Sec.~\ref{sec:transport} is incorporating an arbitrary number of spin components so as to allow simulating transport through a multiterminal quantum dot.
The number of spin components can be controlled in ultracold atom experiments by selectively loading particular hyperfine states.
Such simple versatility is one advantage of our spin-space transport, which serves as a complement to the delicately designed real-space transport~\cite{Krinner:2017}.
The other direction studied in Sec.~\ref{sec:interaction} is incorporating a weak interparticle interaction between two spin components of fermions.
This is particularly relevant to ultracold atom experiments because, unlike quantum dot experiments where the left and right leads are spatially separated, the two spin superposition states occupy the same space so as to interact with each other via a short-range potential.
Finally, our conclusions are summarized in Sec.~\ref{sec:conclusion}.

\section{Quantum transport}\label{sec:transport}
We first describe the scheme to simulate quantum transport with ultracold atoms by employing spin superpositions of fermions interacting with spin-dependent potentials~\cite{Knap:2012,Nishida:2016,You:2019}.
In particular, we extend this scheme to an arbitrary number of spin components so as to allow simulating transport through a multiterminal quantum dot~\cite{Buttiker:1986,Buttiker:1988}.

\subsection{Multiple channels}
To set notations employed below, we consider $N$-component fermions in $d$ spatial dimensions, whose annihilation and creation operators satisfy $\{\psi_\sigma(\x),\psi_\tau^\+(\y)\}=\delta_{\sigma\tau}\delta(\x-\y)$.
We also introduce annihilation and creation operators on a different basis related by a unitary transformation,
\begin{align}
\psi_\alpha(\x) = \sum_\sigma U_{\alpha\sigma}\psi_\sigma(\x), \quad
\psi_\beta^\+(\y) = \sum_\tau\psi_\tau^\+(\y)U_{\tau\beta}^\+,
\end{align}
which satisfy the same anticommutation relation.
We assume that $\psi_\sigma$ diagonalizes the interaction potential matrix and refer to its index $\sigma=1,2,\dots,N$ as {\em spin.}
On the other hand, we choose $\psi_\alpha$ to diagonalize the chemical potential matrix and refer to its index $\alpha=1,2,\dots,N$ as {\em channel.}
Quantities on different bases are to be distinguished by their indices.
The second-quantized Hamiltonian on the spin basis then reads
\begin{align}
H = \sum_\sigma\int\!d\x\,\psi_\sigma^\+(\x)
\left[-\frac{\grad^2}{2m} + V_\sigma(\x)\right]\psi_\sigma(\x),
\end{align}
where $m$ is the mass of a fermion and the spin-dependent single-particle potential $V_\sigma(\x)$ created either by an immobile atom or by an external field can be turned on and off at will.
The interparticle interaction is neglected in this section and we set $\hbar=k_B=1$ throughout this paper.

We suppose that the single-particle potential is initially turned off and apply spin rotations of $|\sigma\>\to|\alpha\>=\sum_\sigma|\sigma\>U_{\sigma\alpha}^\+$ transforming each spin state into some superposition state, which can be performed by coupling two spin states with a resonant laser field.
By successively coupling different pairs of spin states, an $N\times N$ unitary matrix is generated as a product of $2\times2$ unitary matrices whose elements are controlled by the Rabi frequency and the duration of applying the resonant laser field [see Eq.~(\ref{eq:unitary}) below].
The system is then prepared at a thermodynamic equilibrium with temperature $T$ and chemical potential $\mu_\alpha$ for each channel,
\begin{align}
\<\tilde\psi_\alpha^\+(\p)\tilde\psi_\beta(\q)\>_0
= \delta_{\alpha\beta}\delta_{\p\q}f_T(\ep-\mu_\alpha),
\end{align}
where $\tilde\psi_\alpha(\p)=L^{-d/2}\int d\x\,e^{-i\p\cdot\x}\,\psi_\alpha(\x)$ is the Fourier transform in a periodic box of linear size $L$, $\ep=\p^2/2m$ is the energy of a free fermion, and $f_T(\eps)=1/(e^{\eps/T}+1)$ is the Fermi-Dirac distribution function.
We finally turn on the single-particle potential, so that the system is now governed by
\begin{align}\label{eq:hamiltonian}
H = \sum_{\alpha,\beta}\int\!d\x\,\psi_\alpha^\+(\x)
\left[-\frac{\grad^2}{2m}\,\delta_{\alpha\beta}
+ V_{\alpha\beta}(\x)\right]\psi_\beta(\x),
\end{align}
with $V_{\alpha\beta}(\x)\equiv\sum_\sigma U_{\alpha\sigma}V_\sigma(\x)U_{\sigma\beta}^\+$.
Because the single-particle potential on the channel basis generally has off-diagonal elements, it causes transports of particle numbers between different channels.

After a sufficiently long time, the system reaches a steady state, where the exact formula for the transported particle number per unit time, i.e., current, can be derived.
According to the scattering theory in quantum mechanics~\cite{Hewson}, the transition rate from one state $|i\>$ to another $|f\>$ is provided by
\begin{align}\label{eq:probability}
\frac{dP_{i\to f}}{dt} = |\<f|\hat\T|i\>|^2\,2\pi\delta(\eps_i-\eps_f).
\end{align}
Here $\hat\T$ is the transition operator, which at scattering energy $\eps$ satisfies
\begin{align}
\hat\T = \hat\V + \hat\V\frac1{\eps-\hat\H_0+i0^+}\hat\T,
\end{align}
with the single-particle Hamiltonian decomposed into $\hat\H=\hat\H_0+\hat\V$.%
\footnote{If $\hat\T$ is replaced by $\hat\V$ in the first-order Born approximation, Eq.~(\ref{eq:probability}) is reduced to Fermi's golden rule.}
Therefore, by taking into account the occupation of each momentum state as well as the reverse transition process, the net current flowing from one channel $\alpha$ to another $\beta$ is expressed by
\begin{align}\label{eq:transition}
I_{\alpha\to\beta}
&= \sum_{\p,\q}|\<\beta\q|\hat\T|\alpha\p\>|^2\,2\pi\delta(\ep-\eq) \notag\\
&\quad \times [f_T(\ep-\mu_\alpha) - f_T(\eq-\mu_\beta)].
\end{align}
Needless to say, a positive current flows from majority to minority channels under the population imbalance.

Because both the kinetic and the potential energy operators are diagonal on the spin basis, the transition operator is also diagonalized by
\begin{align}
\<\beta\q|\hat\T|\alpha\p\>
= \sum_\sigma U_{\beta\sigma}\<\sigma\q|\hat\T|\sigma\p\>U_{\sigma\alpha}^\+.
\end{align}
In particular, when the single-particle potential is isotropic, $V_\sigma(\x)=V_\sigma(|\x|)$, the matrix element of $\<\sigma\q|\hat\T|\sigma\p\>$ depends only on the scattering energy $\ep$ and the relative angle between the incoming and outgoing momenta, $\cos\chi_{\p\q}\equiv\p\cdot\q/|\p||\q|$, so that it can be expanded as
\begin{align}
& L^d\<\sigma\q|\hat\T|\sigma\p\> \notag\\ &=
\begin{dcases}
\sum_{\ell=0,1}(\cos\chi_{\p\q})^\ell\T_\sigma^\ell(\ep) & (d=1) \\
\sum_{\ell=0,1,2,\dots}(2-\delta_{\ell0})\cos(\ell\chi_{\p\q})\T_\sigma^\ell(\ep) & (d=2) \\
\sum_{\ell=0,1,2,\dots}(2\ell+1)P_\ell(\cos\chi_{\p\q})\T_\sigma^\ell(\ep) & (d=3).
\end{dcases}
\end{align}
Here $\ell=0,1$ for $d=1$ refer to even and odd parity, respectively, $\ell\in\mathbb{N}$ for $d=2,3$ is an orbital angular momentum, and the transition matrix element in each partial-wave sector is related to the potential phase shift according to~\cite{Hewson}
\begin{align}
\T_\sigma^\ell(\ep) =
\begin{dcases}
\frac{i|\p|}{2m}\left[e^{2i\delta_\sigma^\ell(\ep)}-1\right] & (d=1) \\
\frac{i}{m}\left[e^{2i\delta_\sigma^\ell(\ep)}-1\right] & (d=2) \\
\frac{i\pi}{m|\p|}\left[e^{2i\delta_\sigma^\ell(\ep)}-1\right] & (d=3).
\end{dcases}
\end{align}
By substituting the resulting expression of $\<\beta\q|\hat\T|\alpha\p\>$ into Eq.~(\ref{eq:transition}), the current in the infinite volume limit is found to be
\begin{align}\label{eq:current}
I_{\alpha\to\beta}
&= \sum_\ell\frac{\N_\ell}{2\pi}\int_0^\infty\!d\eps\,\biggl|\sum_\sigma
U_{\beta\sigma}e^{2i\delta_\sigma^\ell(\eps)}U_{\sigma\alpha}^\+\biggr|^2 \notag\\
&\quad \times [f_T(\eps-\mu_\alpha) - f_T(\eps-\mu_\beta)],
\end{align}
where the orbital degeneracy factor reads $\N_\ell=1$ for $d=1$, $\N_\ell=2-\delta_{\ell0}$ for $d=2$, and $\N_\ell=2\ell+1$ for $d=3$.
This formula corresponds to the current through a multiterminal quantum dot~\cite{Datta}.
The characteristics of ``quantum dot'' in our scheme are controlled by the spin rotation matrix as well as the spin-dependent single-particle potential via its phase shifts.%
\footnote{From the solution to the radial Schr\"odinger equation,
$$\left[-\frac{d^2}{dr^2}-\frac{d-1}{r}\frac{d}{dr}+\frac{\ell(\ell+d-2)}{r^2}+2mV_\sigma(r)\right]\Psi(r)=k^2\Psi(r),$$
satisfying $\lim_{r\to0}\Psi(r)\propto r^l$, the phase shift at $\eps_k=k^2/2m$ is extracted according to
$$\tan\delta_\sigma^\ell(\eps_k)=\lim_{r\to\infty}\frac{k\J_\ell'(kr)-\J_\ell(kr)\Psi'(r)/\Psi(r)}{k\Y_\ell'(kr)-\Y_\ell(kr)\Psi'(r)/\Psi(r)},$$
where $\J_\ell(x)\equiv x^{1-d/2}J_{\ell-1+d/2}(x)$ and $\Y_\ell(x)\equiv x^{1-d/2}Y_{\ell-1+d/2}(x)$ are defined in terms of the Bessel functions of first and second kinds.}

\subsection{Two channels}
If we specialize to the case of $N=2$, the current formula can be further simplified.
Because the most general $2\times2$ unitary matrix is
\begin{align}\label{eq:unitary}
U = e^{i\zeta}
\begin{pmatrix}
e^{i\eta}\cos\frac\vartheta2 & e^{i\varphi}\sin\frac\vartheta2 \\
-e^{-i\varphi}\sin\frac\vartheta2 & e^{-i\eta}\cos\frac\vartheta2
\end{pmatrix},
\end{align}
Eq.~(\ref{eq:current}) is reduced to
\begin{align}
I_{+\to-}
&= \sum_\ell\frac{\N_\ell}{2\pi}\int_0^\infty\!d\eps\,\sin^2\vartheta\,
\sin^2[\delta_\up^\ell(\eps)-\delta_\down^\ell(\eps)] \notag\\
&\quad \times [f_T(\eps-\mu_+) - f_T(\eps-\mu_-)],
\end{align}
where the spin and channel indices are labeled by $\sigma={\up},\down$ and $\alpha=+,-$, respectively, and the latter play the role of left and right leads in quantum dot experiments.
Therefore, ``quantum dot'' in our scheme is now characterized only by the superposition weight and the difference between the two phase shifts.
Here $\vartheta=\Omega t$ can be controlled by the Rabi frequency $\Omega$ and the duration $t$ of applying a resonant laser field coupling the two spin states and the transport is optimized for $\vartheta=\pi/2$ corresponding to equally weighted superpositions.

\begin{figure}[t]
\includegraphics[width=0.9\columnwidth]{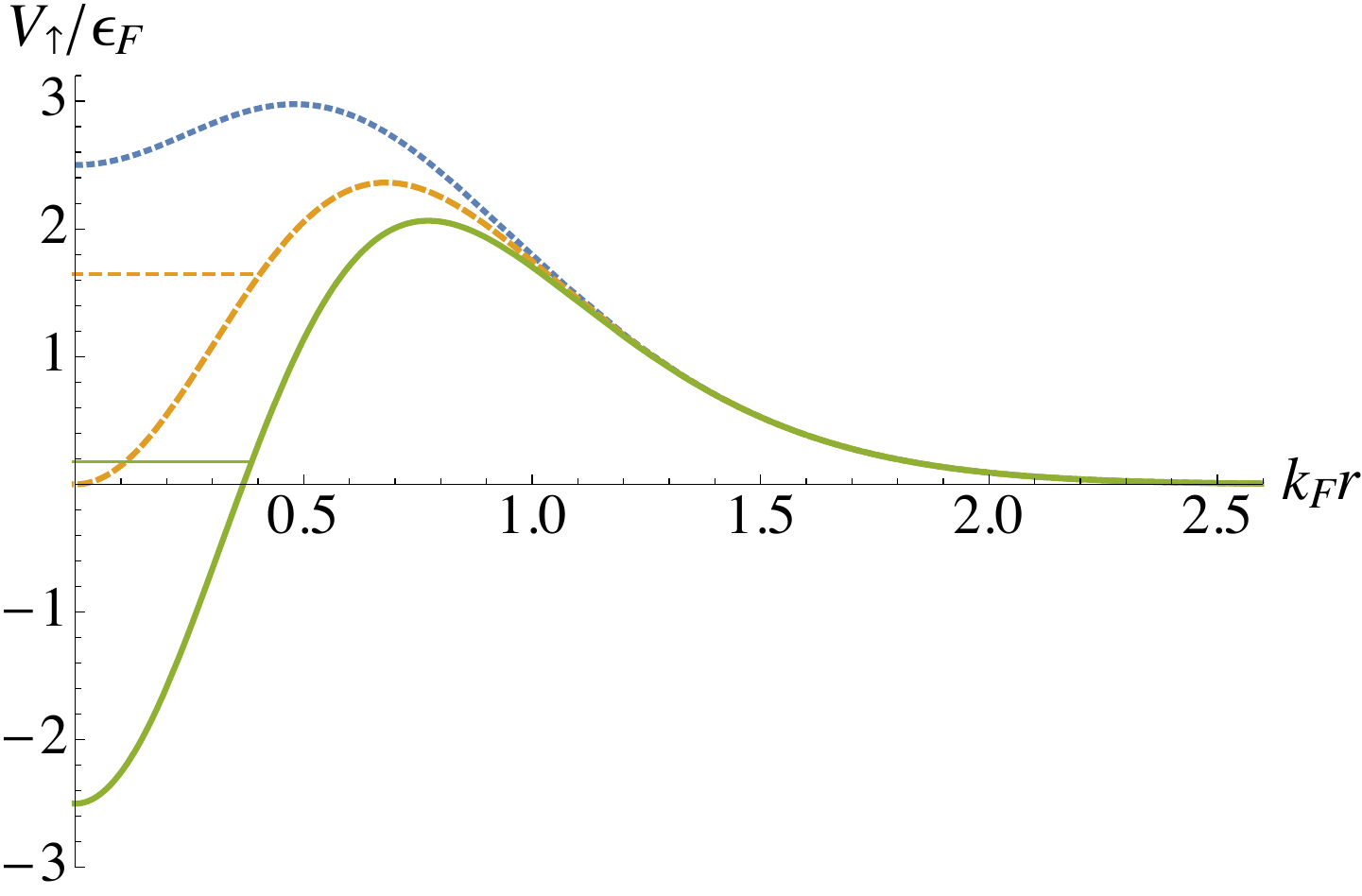}\bigskip\\
\includegraphics[width=0.9\columnwidth]{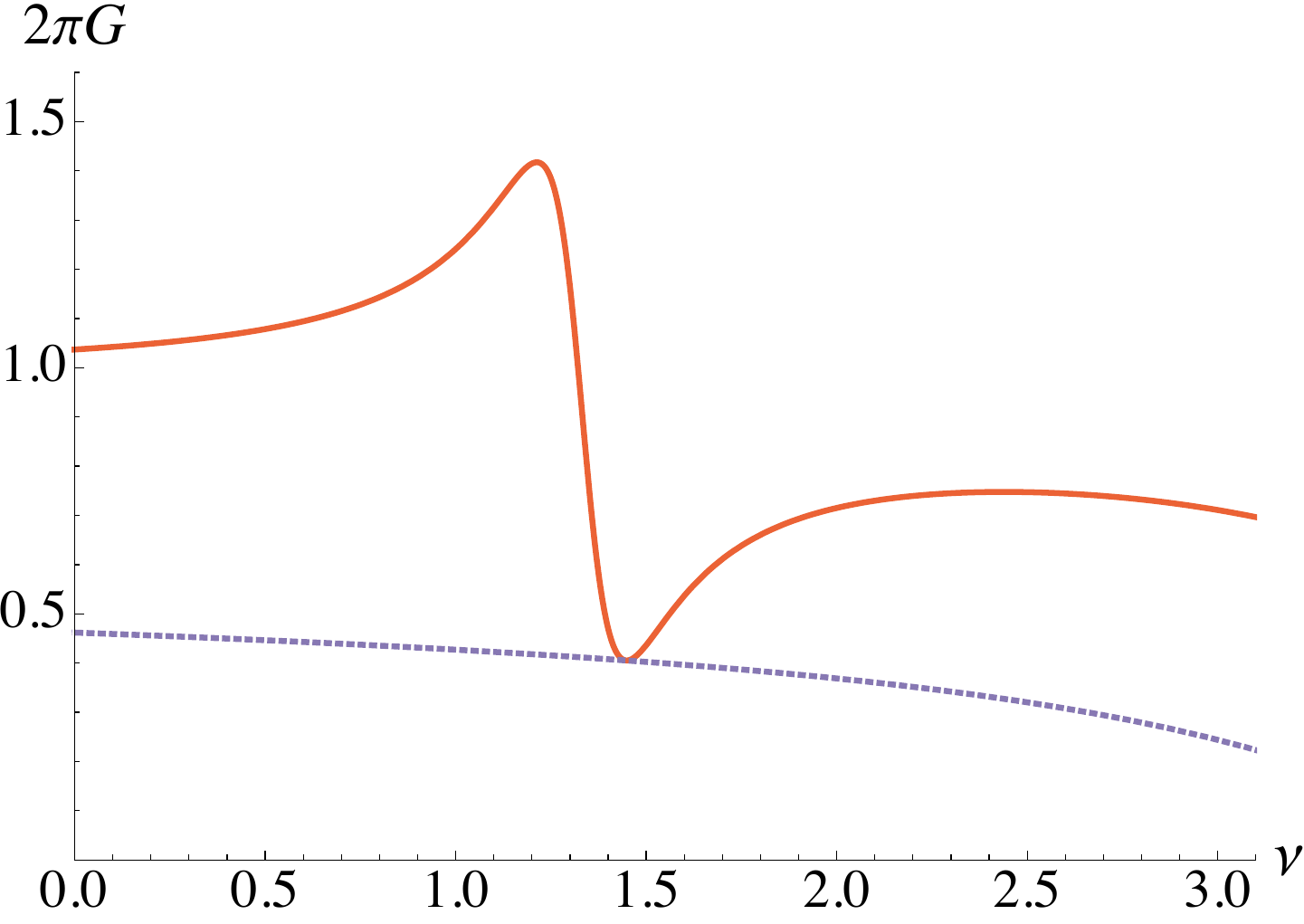}
\caption{\label{fig:conductance}
Upper panel: Model potential $V_\up(r)$ in Eq.~(\ref{eq:model}) for $\nu=0.5$ (dotted), $\nu=1.0$ (dashed), and $\nu=1.5$ (solid).
The horizontal lines indicate resonance energies at which $\delta_\up^\ell(\eps)=\pi/2$ (mod $\pi$) is crossed for $\ell=0$.
Lower panel: Zero-temperature linear conductance $G$ in Eq.~(\ref{eq:conductance}) multiplied by $2\pi$ for $d=1$ with $\vartheta=\pi/2$ as a function of $\nu$.
The dotted curve indicates the nonresonant contribution from $\ell=1$.}
\end{figure}

The linear conductance at zero temperature is then provided by
\begin{align}\label{eq:conductance}
G &\equiv \lim_{T,\Delta\mu\to0}\frac{I_{+\to-}}{\Delta\mu} \notag\\
&= \sum_\ell\frac{\N_\ell}{2\pi}\sin^2\vartheta\,
\sin^2[\delta_\up^\ell(\eF)-\delta_\down^\ell(\eF)],
\end{align}
where $\eF=\kF^2/2m>0$ is the Fermi energy and $\Delta\mu$ is the chemical potential difference with $\mu_\pm=\eF\pm\Delta\mu/2$.
In ultracold atom experiments, the transport actually relaxes the population imbalance toward zero.
Within a quasi\-steady approximation assuming the system to remain equilibrated at every moment during the slow transport, the linear conductance is measurable by monitoring the exponential decrease of
\begin{align}\label{eq:measurement}
\Delta N(t) = e^{-(2G/\kappa)t}\,\Delta N(0),
\end{align}
where $\Delta N=N_+-N_-$ is the particle number difference and $\kappa\equiv\Delta N/\Delta\mu=-\sum_\p f_T'(\ep-\eF)$ is the compressibility~\cite{Krinner:2017}.
As a simple demonstration, we consider single-particle potentials of $V_\down(r)=0$ and
\begin{align}\label{eq:model}
V_\up(r) = 5\eF\left[e^{-(\kF r)^2} - \nu\,e^{-(2\kF r)^2}\right],
\end{align}
which consisting of an outer repulsive barrier and an inner attractive well models a quantum dot and can be created by superimposing two focused laser fields.
The linear conductance at zero temperature is then computed for $d=1$ with $\vartheta=\pi/2$ as shown in Fig.~\ref{fig:conductance}.
Here a Fano resonance is found with increasing $\nu$ so that the resonance energy crosses the Fermi energy ($\nu\approx1.2$), where the conductance per each partial wave reaches the maximal value of $G_\ell=\N_\ell/2\pi$ allowed by the unitarity.
We note that the same Fano resonance can also be found for $d=2$ and 3, while the nonresonant contribution from the other partial-wave sectors tends to be large in total.

Another way to create a single-particle potential in ultracold atom experiments is by an immobile atom.
By tuning its interaction with spin-up fermions via a magnetic-field-induced Feshbach resonance, the resulting single-particle potential is modeled by a zero-range potential with a scattering length $a$.
Its phase shift for $\ell=0$ is provided by
\begin{align}
\cot\delta_\up^\ell(\eps_k) =
\begin{dcases}
ak & (d=1) \\
\frac2\pi\ln ak & (d=2) \\
-\frac1{ak} & (d=3),
\end{dcases}
\end{align}
while $\delta_\up^\ell(\eps)=0$ for all $\ell\geq1$.%
\footnote{Here the normalization of $a$ is chosen so that a bound state appears at $\eps=-1/(2ma^2)$ for $a>0$~\cite{Fujii:2018}.}
Therefore, the unitarity-limited conductance of $G_{\ell=0}=1/2\pi$ is reached at $a\kF=\pm0$, 1 and $\infty$ for $d=1$, 2, and 3, respectively.
If a finite number of such immobile atoms is present in the system, the total current is simply multiplied by their number as long as they are far separated so as to be uncorrelated.

\section{Interaction effects}\label{sec:interaction}
Unlike quantum dot experiments where the left and right leads are spatially separated, fermions of different channels in our scheme occupy the same space so as to interact with each other via a short-range potential:%
\footnote{This interaction Hamiltonian invariant under spin rotation takes the same form both on the spin and the channel bases and does not cause undesired bulk transports between different channels.}
\begin{align}\label{eq:interaction}
H_\mathrm{int} = \frac{g}{2}\sum_{\alpha,\beta}\int\!d\x\,\psi_\alpha^\+(\x)
\psi_\beta^\+(\x)\psi_\beta(\x)\psi_\alpha(\x).
\end{align}
A weak interparticle interaction for $d=3$ is negligible in the dilute limit because it is irrelevant, while it may not be the case in lower dimensions.
Here we study how a weak interparticle interaction in bulk affects quantum transport for $d=1$ in the case of $N=2$, which is facilitated by employing the bosonization and renormalization techniques~\cite{Kane:1992a,Kane:1992b}.

\subsection{Bosonization}
Low-energy physics in one dimension is dominated by excitations about the left and right Fermi points.
With the fermion annihilation operator expanded as
\begin{align}
\psi_\alpha(x) \simeq e^{-i\kF x}\psi_{\alpha L}(x) + e^{i\kF x}\psi_{\alpha R}(x),
\end{align}
the local potential term of Eq.~(\ref{eq:hamiltonian}) is decomposed into
\begin{align}\label{eq:potential}
H_\mathrm{pot} &\simeq \sum_\alpha u_\alpha
\left[\psi_{\alpha L}^\+(0)\psi_{\alpha L}(0)
+ \psi_{\alpha R}^\+(0)\psi_{\alpha R}(0)\right] \notag\\
&\quad + \sum_\alpha\bar u_\alpha
\left[\psi_{\alpha L}^\+(0)\psi_{\bar\alpha L}(0)
+ \psi_{\alpha R}^\+(0)\psi_{\bar\alpha R}(0)\right] \notag\\
&\quad + \sum_\alpha w_\alpha\,
\psi_{\alpha R}^\+(0)\psi_{\alpha L}(0) + \mathrm{H.c.} \notag\\
&\quad + \sum_\alpha\bar w_\alpha\,
\psi_{\alpha R}^\+(0)\psi_{\bar\alpha L}(0) + \mathrm{H.c.}.
\end{align}
Here the four kinds of terms with $\bar\alpha\equiv-\alpha$ correspond to forward ($u_\alpha$, $\bar u_\alpha$) and backward ($w_\alpha$, $\bar w_\alpha$) scatterings without ($u_\alpha$, $w_\alpha$) and with ($\bar u_\alpha$, $\bar w_\alpha$) channel transitions.
The coupling of each scattering process is provided by $u_\alpha=\tilde V_{\alpha\alpha}(0)$, $\bar u_\alpha=\tilde V_{\alpha\bar\alpha}(0)$, $w_\alpha=\tilde V_{\alpha\alpha}(2\kF)$, and $\bar w_\alpha=\tilde V_{\alpha\bar\alpha}(2\kF)$, where $\tilde V_{\alpha\beta}(p)=\int dx\,e^{-ipx}\,V_{\alpha\beta}(|x|)$ is the Fourier transform of the single-particle potential.
In particular, it is $\bar u_\alpha$ and $\bar w_\alpha$ that cause transports of particle numbers between different channels.
We note that, if the interparticle interaction was absent, the zero-temperature linear conductance to their lowest order in perturbation would be
\begin{align}\label{eq:free}
G = \frac{|\bar u_\alpha|^2 + |\bar w_\alpha|^2}{\pi\vF^2} \qquad (g=0),
\end{align}
which readily follows from Eq.~(\ref{eq:transition}) with $\vF=\kF/m$ being the Fermi velocity.%
\footnote{Because of $\bar u_\alpha^*=\bar u_{\bar\alpha}$ and $\bar w_\alpha^*=\bar w_{\bar\alpha}$, $|\bar u_\alpha|^2$ and $|\bar w_\alpha|^2$ are actually independent of $\alpha=\pm$.}

On the other hand, the standard bosonization formula~\cite{Giamarchi},
\begin{subequations}\label{eq:bosonization}
\begin{align}
\psi_{\pm L} &= \frac{F_{\pm L}}{\sqrt{2\pi\lambda}}\,
e^{i\{[\theta_c(x)+\phi_c(x)]\pm[\theta_s(x)+\phi_s(x)]\}/\sqrt2}, \\
\psi_{\pm R} &= \frac{F_{\pm R}}{\sqrt{2\pi\lambda}}\,
e^{i\{[\theta_c(x)-\phi_c(x)]\pm[\theta_s(x)-\phi_s(x)]\}/\sqrt2},
\end{align}
\end{subequations}
brings the bulk Hamiltonian into $H_0+H_\mathrm{int}\simeq H_c+H_s$ so as to separate the charge sector,
\begin{align}
H_c = \frac{v_c}{2\pi}\int\!dx\left\{K_c[\nabla\theta_c(x)]^2
+ \frac1{K_c}[\nabla\phi_c(x)]^2\right\},
\end{align}
and the spin sector,%
\footnote{Although common wording of ``spin'' is employed here and below, we work on the channel basis throughout this section.}
\begin{align}\label{eq:spin}
H_s &= \frac{v_s}{2\pi}\int\!dx\left\{K_s[\nabla\theta_s(x)]^2
+ \frac1{K_s}[\nabla\phi_s(x)]^2\right\} \notag\\
&\quad + \frac{2g}{(2\pi a)^2}\int\!dx\,\cos[\sqrt8\,\phi_s(x)].
\end{align}
Here $\theta_{c,s}$ and $\phi_{c,s}$ are the boson fields associated with phase and density fluctuations, respectively, $F_{\pm L,R}$ is the Klein factor, and $\lambda\sim\kF^{-1}$ is a short-distance cutoff scale, while the charge or spin velocity and Tomonaga-Luttinger parameter are provided by $v_{c,s}=\vF\sqrt{1\pm g/\pi\vF}$ and $K_{c,s}=1/\sqrt{1\pm g/\pi\vF}$ (upper sign for $c$ and lower sign for $s$), respectively.

\subsection{Renormalization}
Because the local potential Hamiltonian does not affect the bulk properties, the renormalization of the sine-Gordon Hamiltonian remains the same~\cite{Giamarchi}.
In particular, for an attractive interaction $g<0$ with $K_s<1$, the cosine term of Eq.~(\ref{eq:spin}) is relevant and thus opens up a spin excitation gap of $\Delta_s\propto\eF e^{\pi\vF/g}$ at weak coupling~\cite{Fuchs:2004}.
Consequently, the transport between different channels is suppressed,
\begin{align}\label{eq:attractive}
G \to 0 \qquad (g<0),
\end{align}
in the low-energy limit $T,\Delta\mu\ll\Delta_s$ because the bulk system turns into a spin insulator.

On the other hand, for a repulsive interaction $g>0$ with $K_s>1$, the cosine term of Eq.~(\ref{eq:spin}) is irrelevant and $g$ is thus renormalized toward zero in the low-energy limit, where a fixed point of $K_s\to K_s^*=1$ is reached for interactions invariant under spin rotation~\cite{Giamarchi}.
Therefore, a spin excitation remains gapless so as to allow for the transport between different channels.

\begin{figure}[t]
\includegraphics[width=0.95\columnwidth]{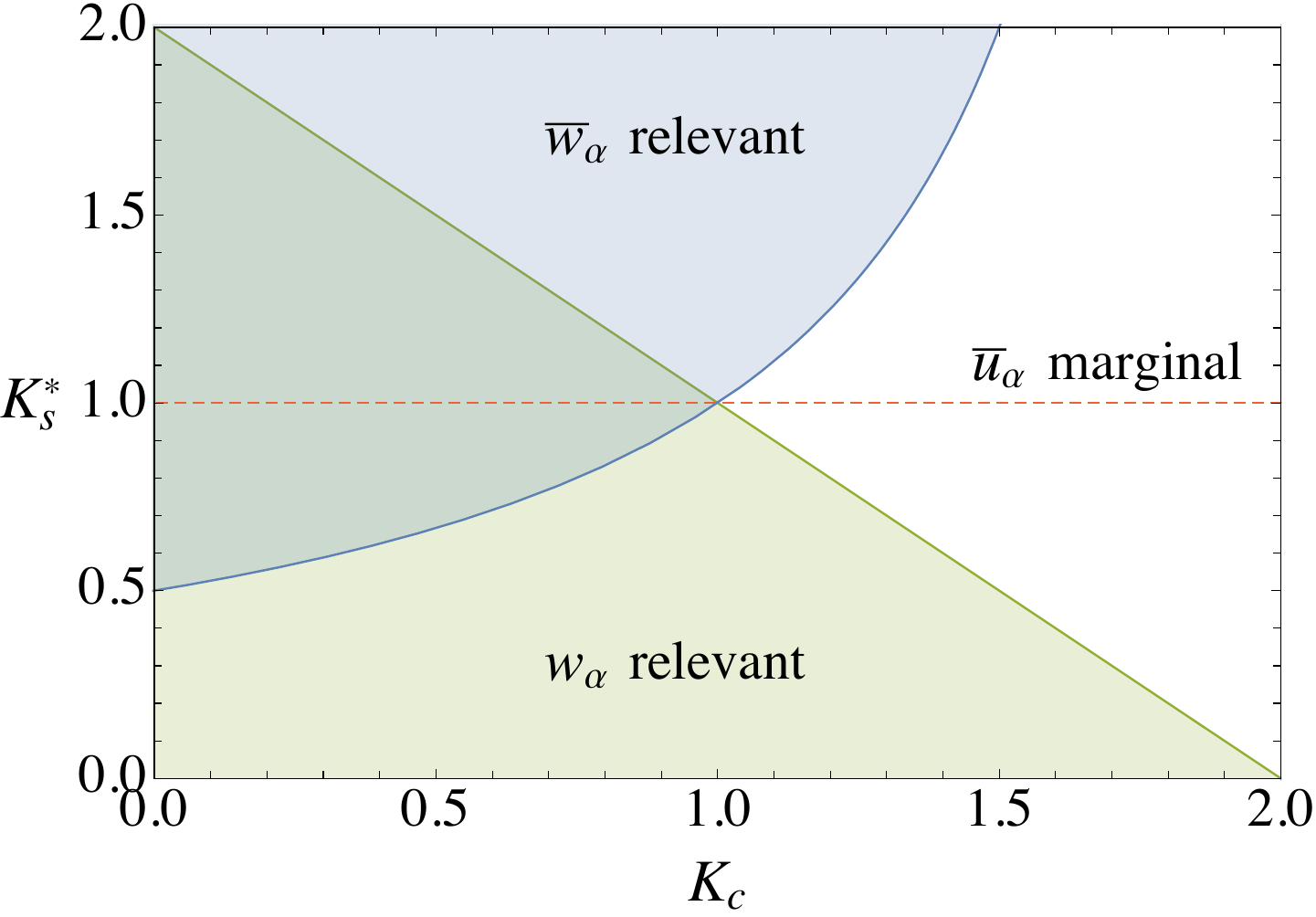}
\caption{\label{fig:coupling}
``Phase diagram'' in the plane of $K_c$ and $K_s^*$ where the upper (lower) shaded region indicates $\bar w_\alpha$ ($w_\alpha$) being relevant, while it is irrelevant elsewhere.
$\bar u_\alpha$ is marginal on the horizontal dashed line and irrelevant elsewhere.}
\end{figure}

To gain further insight, we then study the renormalization of the local potential Hamiltonian.
By denoting each fermion bilinear operator in Eq.~(\ref{eq:potential}) as $\O(0)$, its scaling dimension $\Delta_\O$ is extracted from the correlation function according to
\begin{align}
\<\O^\+(x)\O(0)\> \propto \frac1{\lambda^2}\left(\frac{\lambda^2}{x^2}\right)^{\Delta_\O}.
\end{align}
Such correlation functions with respect to the bulk Hamiltonian of $H_c+H_s|_{g\to0,K_s\to K_s^*}$ at the fixed point can be computed by bosonizing $\O$ with Eq.~(\ref{eq:bosonization}) as detailed in Ref.~\cite{Giamarchi}.
Consequently, $\Delta_\O=0$, $(K_s^*+1/K_s^*)/2$, $(K_c+K_s^*)/2$, and $(K_c+1/K_s^*)/2$ are found for $\O=\psi_{\alpha L}^\+\psi_{\alpha L}$, $\psi_{\alpha L}^\+\psi_{\bar\alpha L}$, $\psi_{\alpha R}^\+\psi_{\alpha L}$, and $\psi_{\alpha R}^\+\psi_{\bar\alpha L}$, respectively.
Because each coupling has a dimension of $1-\Delta_\O$, its renormalization group equation up to the linear order is provided by
\begin{align}
\frac{d u_\alpha}{d l} &= u_\alpha, \\
\frac{d\bar u_\alpha}{d l}
&= \frac12\left(2-K_s^*-\frac1{K_s^*}\right)\bar u_\alpha, \\
\frac{d w_\alpha}{d l} &= \frac12\left(2-K_c-K_s^*\right)w_\alpha, \\
\frac{d\bar w_\alpha}{d l}
&= \frac12\left(2-K_c-\frac1{K_s^*}\right)\bar w_\alpha.
\end{align}
Therefore, we find that $u_\alpha$ is relevant, $\bar u_\alpha$ is marginal for $K_s^*=1$ but otherwise irrelevant, and $w_\alpha$ and $\bar w_\alpha$ are relevant for the parameter regions indicated in Fig.~\ref{fig:coupling}.

In particular, $\bar w_\alpha$ responsible for quantum transport is found to be relevant for a repulsive interaction with $K_c=1-g/2\pi\vF<1$ and $K_s^*=1$, so that it grows toward the low-energy limit as
\begin{align}
\bar w_\alpha(l) = e^{(g/4\pi\vF)l}\,\bar w_\alpha.
\end{align}
By running the renormalization down to the scale of temperature, $T=e^{-l}\eF$, the linear conductance is dominated by
\begin{align}\label{eq:repulsive}
G \propto \frac{|\bar w_\alpha(l)|^2}{\vF^2}
\to \left(\frac\eF{T}\right)^{g/2\pi\vF}\frac{|\bar w_\alpha|^2}{\vF^2} \qquad (g>0),
\end{align}
which is enhanced by the power law with lowering the temperature in the range of $\Delta\mu\ll T\ll\eF$.
On the other hand, $T$ in Eq.~(\ref{eq:repulsive}) is to be replaced by $\Delta\mu$ in the range of $T\ll\Delta\mu\ll\eF$, where the zero-temperature conductance in turn exhibits the nonlinear current-voltage characteristic.
Such stimulated transport may be understood because the bulk spin-density-wave quasiorder tends to locally increase the population imbalance between different channels.

\section{Conclusions}\label{sec:conclusion}
In this paper, we studied quantum transport simulated with ultracold atoms by employing spin superpositions of fermions interacting with spin-dependent potentials~\cite{Knap:2012,Nishida:2016,You:2019}.
This scheme was first extended to an arbitrary number of spin components so as to allow simulating transport through a multiterminal quantum dot, where the current formula was derived in terms of a spin rotation matrix and potential phase shifts [Eq.~(\ref{eq:current})].
We then showed that a Fano resonance manifests itself in the case of two spin components in measuring a linear conductance at zero temperature with deforming the single-particle potential so that the resonance energy crosses the Fermi energy (Fig.~\ref{fig:conductance}).

We also studied how a weak interparticle interaction in bulk affects quantum transport in one dimension with the bosonization and renormalization techniques.
Depending on whether the interparticle interaction is attractive, vanishing, or repulsive, the conductance in the low-energy limit $T,\Delta\mu\ll\eF$ was found to exhibit the three distinct behaviors of
\begin{align}
G \to
\begin{dcases}
0 & (g<0) \\
\mathrm{const} & (g=0) \\
\max(T,\Delta\mu)^{-g/2\pi\vF} & (g>0),
\end{dcases}
\end{align}
according to Eqs.~(\ref{eq:attractive}), (\ref{eq:free}), and (\ref{eq:repulsive}), respectively.
Therefore, while the conductance vanishes for an attractive interaction due to the bulk spin gap, it is enhanced for a repulsive interaction by the power law with lowering the temperature or the chemical potential difference.
Because the linear conductance is measurable in ultracold atom experiments [Eq.~(\ref{eq:measurement})], our findings here are hopefully to be observed in future experiments.

\acknowledgments
The authors thank Professor Yoshiro Takahashi and his group members for valuable discussions.
This work was supported by JSPS KAKENHI Grants No.\ JP17K14366 and No.\ JP18H05405.
One of the authors (S.U.) was also supported by Matsuo Foundation and Waseda University Grant for Special Research Projects (No.\ 2019C-461).


\begin{thebibliography}{99}

\bibitem{Georgescu:2014}
I.~M.~Georgescu, S.~Ashhab, and F.~Nori,
``Quantum simulation,''
\href{https://doi.org/10.1103/RevModPhys.86.153}
{Rev.\ Mod.\ Phys.\ \textbf{86}, 153-185 (2014)}.

\bibitem{Bloch:2012}
I.~Bloch, J.~Dalibard, and S.~Nascimb\`ene,
``Quantum simulations with ultracold quantum gases,''
\href{https://doi.org/10.1038/nphys2259}
{Nat.\ Phys.\ \textbf{8}, 267-276 (2012)}.

\bibitem{Gross:2017}
C.~Gross and I.~Bloch,
``Quantum simulations with ultracold atoms in optical lattices,''
\href{https://doi.org/10.1126/science.aal3837}
{Science \textbf{357}, 995-1001 (2017)}.

\bibitem{Tarruell:2018}
L.~Tarruell and L.~Sanchez-Palencia,
``Quantum simulation of the Hubbard model with ultracold fermions in optical lattices,''
\href{https://doi.org/10.1016/j.crhy.2018.10.013}
{C.\ R.\ Phys.\ \textbf{19}, 365-393 (2018)}.

\bibitem{Wiese:2013}
U.-J.~Wiese,
``Ultracold quantum gases and lattice systems: Quantum simulation of lattice gauge theories,''
\href{https://doi.org/10.1002/andp.201300104}
{Ann.\ Phys.\ (Berlin) \textbf{525}, 777-796 (2013)}.

\bibitem{Zohar:2016}
E.~Zohar, J.~I.~Cirac, and B.~Reznik,
``Quantum simulations of lattice gauge theories using ultracold atoms in optical lattices,''
\href{https://doi.org/10.1088/0034-4885/79/1/014401}
{Rep.\ Prog.\ Phys.\ \textbf{79}, 014401 (2016)}.

\bibitem{Boada:2012}
O.~Boada, A.~Celi, J.~I.~Latorre, and M.~Lewenstein,
``Quantum simulation of an extra dimension,''
\href{https://doi.org/10.1103/PhysRevLett.108.133001}
{Phys.\ Rev.\ Lett.\ \textbf{108}, 133001 (2012)}.

\bibitem{Celi:2014}
A.~Celi, P.~Massignan, J.~Ruseckas, N.~Goldman, I.~B.~Spielman, G.~Juzeli\=unas, and M.~Lewenstein,
``Synthetic gauge fields in synthetic dimensions,''
\href{https://doi.org/10.1103/PhysRevLett.112.043001}
{Phys.\ Rev.\ Lett.\ \textbf{112}, 043001 (2014)}.

\bibitem{Mancini:2015}
M.~Mancini, G.~Pagano, G.~Cappellini, L.~Livi, M.~Rider, J.~Catani, C.~Sias, P.~Zoller, M.~Inguscio, M.~Dalmonte, and L.~Fallani,
``Observation of chiral edge states with neutral fermions in synthetic Hall ribbons,''
\href{https://doi.org/10.1126/science.aaa8736}
{Science \textbf{349}, 1510-1513 (2015)}.

\bibitem{Stuhl:2015}
B.~K.~Stuhl, H.-I.~Lu, L.~M.~Aycock, D.~Genkina, and I.~B.~Spielman,
``Visualizing edge states with an atomic Bose gas in the quantum Hall regime,''
\href{https://doi.org/10.1126/science.aaa8515}
{Science \textbf{349}, 1514-1518 (2015)}.

\bibitem{Price:2015}
H.~M.~Price, O.~Zilberberg, T.~Ozawa, I.~Carusotto, and N.~Goldman,
``Four-dimensional quantum Hall effect with ultracold atoms,''
\href{https://doi.org/10.1103/PhysRevLett.115.195303}
{Phys.\ Rev.\ Lett.\ \textbf{115}, 195303 (2015)}.

\bibitem{Lee:2018}
C.~H.~Lee, Y.~Wang, Y.~Chen, and X.~Zhang,
``Electromagnetic response of quantum Hall systems in dimensions five and six and beyond,''
\href{https://doi.org/10.1103/PhysRevB.98.094434}
{Phys.\ Rev.\ B \textbf{98}, 094434 (2018)}.

\bibitem{Petrides:2018}
I.~Petrides, H.~M.~Price, and O.~Zilberberg,
``Six-dimensional quantum Hall effect and three-dimensional topological pumps,''
\href{https://doi.org/10.1103/PhysRevB.98.125431}
{Phys.\ Rev.\ B \textbf{98}, 125431 (2018)}.

\bibitem{Ozawa:2019}
T.~Ozawa and H.~M.~Price,
``Topological quantum matter in synthetic dimensions,''
\href{https://doi.org/10.1038/s42254-019-0045-3}
{Nat.\ Rev.\ Phys.\ \textbf{1}, 349-357 (2019)}.

\bibitem{Knap:2012}
M.~Knap, A.~Shashi, Y.~Nishida, A.~Imambekov, D.~A.~Abanin, and E.~Demler,
``Time-dependent impurity in ultracold fermions: Orthogonality catastrophe and beyond,''
\href{https://doi.org/10.1103/PhysRevX.2.041020}
{Phys.\ Rev.\ X \textbf{2}, 041020 (2012)}.

\bibitem{You:2019}
J.-S.~You, R.~Schmidt, D.~A.~Ivanov, M.~Knap, and E.~Demler,
``Atomtronics with a spin: Statistics of spin transport and nonequilibrium orthogonality catastrophe in cold quantum gases,''
\href{https://doi.org/10.1103/PhysRevB.99.214505}
{Phys.\ Rev.\ B \textbf{99}, 214505 (2019)}.

\bibitem{Nishida:2013}
Y.~Nishida,
``SU(3) orbital Kondo effect with ultracold atoms,''
\href{https://doi.org/10.1103/PhysRevLett.111.135301}
{Phys.\ Rev.\ Lett.\ \textbf{111}, 135301 (2013)}.

\bibitem{Nishida:2016}
Y.~Nishida,
``Transport measurement of the orbital Kondo effect with ultracold atoms,''
\href{https://doi.org/10.1103/PhysRevA.93.011606}
{Phys.\ Rev.\ A \textbf{93}, 011606(R) (2016)}.

\bibitem{Krinner:2017}
S.~Krinner, T.~Esslinger, and J.-P.~Brantut,
``Two-terminal transport measurements with cold atoms,''
\href{https://doi.org/10.1088/1361-648X/aa74a1}
{J.\ Phys.: Condens.\ Matter \textbf{29}, 343003 (2017)}.

\bibitem{Buttiker:1986}
M.~B\"uttiker,
``Four-terminal phase-coherent conductance,''
\href{https://doi.org/10.1103/PhysRevLett.57.1761}
{Phys.\ Rev.\ Lett.\ \textbf{57}, 1761-1764 (1986)}.

\bibitem{Buttiker:1988}
M.~B\"uttiker,
``Absence of backscattering in the quantum Hall effect in multiprobe conductors,''
\href{https://doi.org/10.1103/PhysRevB.38.9375}
{Phys.\ Rev.\ B \textbf{38}, 9375-9389 (1988)}.

\bibitem{Hewson}
A.~C.~Hewson,
\textit{The Kondo Problem to Heavy Fermions}
(Cambridge University Press, Cambridge, UK, 1993).

\bibitem{Datta}
S.~Datta,
\textit{Electronic Transport in Mesoscopic Systems}
(Cambridge University Press, Cambridge, UK, 1995).

\bibitem{Fujii:2018}
K.~Fujii and Y.~Nishida,
``Hydrodynamics with spacetime-dependent scattering length,''
\href{https://doi.org/10.1103/PhysRevA.98.063634}
{Phys.\ Rev.\ A \textbf{98}, 063634 (2018)}.

\bibitem{Kane:1992a}
C.~L.~Kane and M.~P.~A.~Fisher,
``Transport in a one-channel Luttinger liquid,''
\href{https://doi.org/10.1103/PhysRevLett.68.1220}
{Phys.\ Rev.\ Lett.\ \textbf{68}, 1220-1223 (1992)}.

\bibitem{Kane:1992b}
C.~L.~Kane and M.~P.~A.~Fisher,
``Transmission through barriers and resonant tunneling in an interacting one-dimensional electron gas,''
\href{https://doi.org/10.1103/PhysRevB.46.15233}
{Phys.\ Rev.\ B \textbf{46}, 15233-15262 (1992)}.

\bibitem{Giamarchi}
T.~Giamarchi,
\textit{Quantum Physics in One Dimension}
(Oxford University Press, Oxford, UK, 2003).

\bibitem{Fuchs:2004}
J.~N.~Fuchs, A.~Recati, and W.~Zwerger,
``Exactly solvable model of the BCS-BEC crossover,''
\href{https://doi.org/10.1103/PhysRevLett.93.090408}
{Phys.\ Rev.\ Lett.\ \textbf{93}, 090408 (2004)}.

\end{thebibliography}
\end{document}